\documentclass[twocolumn,prl,showpacs,floatfix]{revtex4}
\usepackage{graphicx}
\usepackage{bm}

\begin{document}

\title{\textbf{Orbital degeneracy removed by charge order in triangular
antiferromagnet AgNiO$_2$}}
\author{E. Wawrzy\'{n}ska$^{1}$, R. Coldea$^{1}$, E.M. Wheeler$^{2,3}$ I.I.
Mazin$^{4}$, M.D. Johannes$^{4}$,\\ T. S\"{o}rgel$^{5}$, M.
Jansen$^{5}$, R.M. Ibberson$^{6}$, P.G. Radaelli$^{6}$}
\affiliation{$^1$H.H. Wills Physics Laboratory, University of
Bristol, Tyndall Avenue, Bristol, BS8 1TL, United Kingdom\\
$^2$Clarendon Laboratory, University of Oxford, Parks Road, Oxford
OX1 3PU, United Kingdom $^3$Institute Laue-Langevin, BP 156, 38042
Grenoble Cedex 9, France \\$^4$Code 6393, Naval Research
Laboratory,
Washington, D.C. 20375\\
$^5$Max-Planck Institut f\"{u}r Festk\"{o}rperforschung,
Heisenbergstrasse 1, D-70569 Stuttgart, Germany \\$^6$ISIS
Facility, Rutherford Appleton Laboratory, Chilton, Didcot OX11
0QX, United Kingdom}
%, $^7$Dept. of Physics and
%Astronomy, University College London, Gower Street, London WC1E
%6BT, United Kingdom}
\date{\today }

\begin{abstract} We report a high-resolution neutron diffraction
study on the orbitally-degenerate spin-1/2 hexagonal metallic
antiferromagnet AgNiO$_2$. A structural transition to a tripled
unit cell with expanded and contracted NiO$_6$ octahedra indicates
$\sqrt{3} \times \sqrt{3}$ charge order on the Ni triangular
lattice. This suggests charge order as a possible mechanism of
lifting the orbital degeneracy in the presence of charge
fluctuations, as an alternative to the more usual Jahn-Teller
distortions. A novel magnetic ground state is observed at low
temperatures with the electron-rich $S=1$ Ni sites arranged in
alternating ferromagnetic rows on a triangular lattice, surrounded
by a honeycomb network of non-magnetic and metallic Ni ions. We
also report first-principles band-structure calculations that
explain microscopically the origin of these phenomena.
\end{abstract}

\maketitle

Electronic systems on triangular lattices have attracted wide
interest theoretically and experimentally due to the possible
existence of unconventional ground states stabilized by the
frustrated lattice geometry \cite{ong}. The layered cobaltate
NaCoO$_{2}$ with a triangular lattice of
orbitally-\emph{non}degenerate Co$^{3+}$ ions \cite{jansen} shows
a number of unusual phases upon hole doping \emph{via}
deintercalation of Na, as well as superconductivity upon hydration
at a particular doping \cite{ong}. The addition of orbital
degeneracy creates more instabilities for cooperative order and
the behavior in the presence of metallic conductivity is not well
understood and largely experimentally unexplored. In more
localized systems such as NaNiO$_{2}$ with spin-1/2 Ni$^{3+}$
ions in the $t_{2g}^{6}e_{g}^{1}$ configuration the two-fold
$e_{g}$ orbital degeneracy is lifted by a Jahn-Teller (JT)
structural distortion to a monoclinic structure \cite{NaNiO2},
which splits the $e_g$ band and opens a gap. However it is unclear
if such a mechanism applies in less localized systems where the
local tendency for JT distortions competes with the charge
transfer between sites, allowed by an efficient metallic screening
\cite{RENIO3}. The metallic layered silver nickelates AgNiO$_2$
\cite{Soergel05,Shin93} and Ag$_2$NiO$_2$ \cite{ag2} are ideal
candidates to explore this physics.

Here we report high-resolution structural studies of AgNiO$_2$ and
find that no JT distortions occur but the ideal structure is
distorted to a {\em hexagonal tripled} unit cell with expanded and
contracted NiO$_6$ octahedra. This implies that the orbital
degeneracy of the formally Ni$^{3+}$ sites is lifted by a novel
$\sqrt{3}\times \sqrt{3}$ charge ordering (CO) that occurs via the
charge transfer $3e_{g}^{1}\rightarrow
e_{g}^{2}+e_{g}^{0.5}+e_{g}^{0.5}$, which to the best of our
knowledge is the first experimental example of such a CO pattern
in a triangular system. Using band-structure calculations we show
that this CO is a 2D analogue of the metal-insulator transition in
the 3D $R$NiO$_{3}$ \cite{RENIO3_exp} attributed to the charge
transfer $2e_{g}^{1}\rightarrow e_{g}^{2}+e_{g}^{0}$
\cite{RENIO3}. However, in contrast to the 3D nickelates in
AgNiO$_2$ the triangular lattice geometry allows the system to
remain metallic in the CO phase, as itinerant electrons can hop on
the inter-connected honeycombe network formed by the two
electron-depleted sublattices ($e_{g}^{0.5}$); those sites remain
magnetically unordered at base temperature. The remaining one
third of Ni sites are strongly magnetic ($e_g^2$) and form an
effectively tripled antiferromagnetic (AFM) triangular lattice,
which orders in an unusual collinear $2\times1$ structure of
alternating stripes.
%This implies
%significant interactions beyond nearest-neighbor AFM
%superexchange.
%The possible r\^{o}le of anisotropy and inter-layer couplings in
%selecting this ordered pattern is proposed.

\begin{figure}[tbhp]
\begin{center}
  \includegraphics[width=7cm,bbllx=71,bblly=429,bburx=504,
  bbury=684,angle=0,clip=]{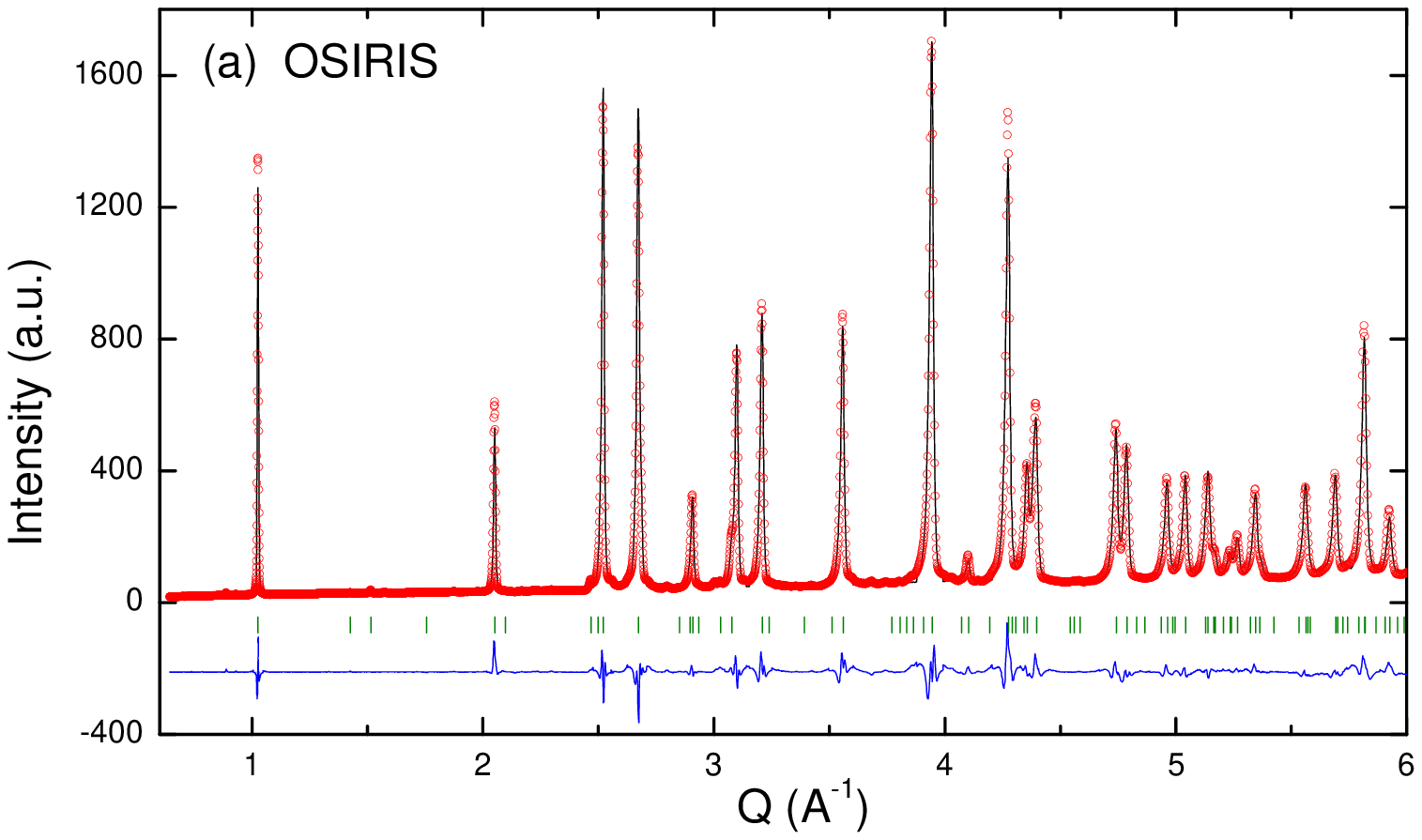}
 \includegraphics[width=7cm,bbllx=103,bblly=219,bburx=555,
  bbury=556,angle=0,clip=]{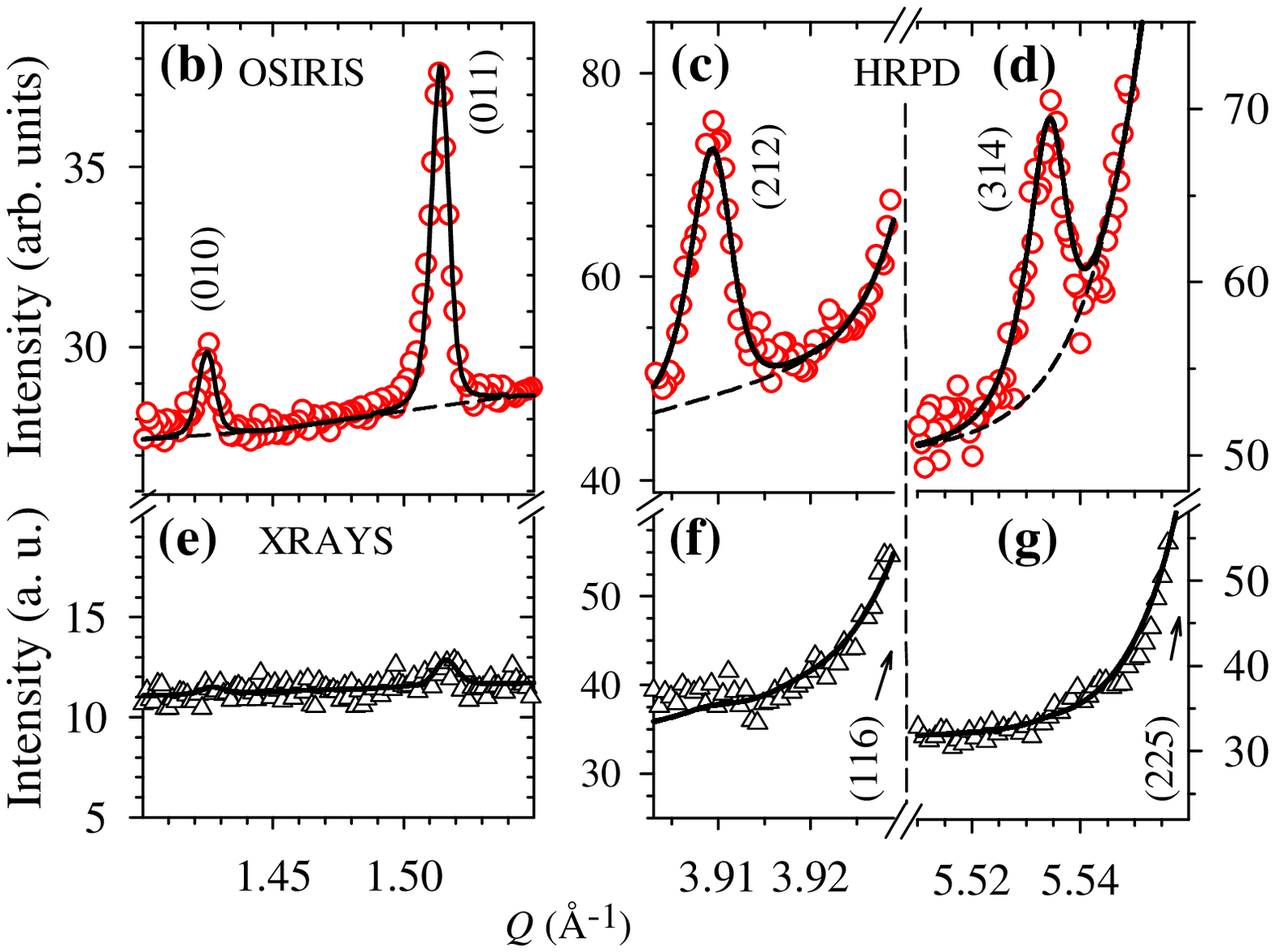}
\caption{\label{fig_pattern} a) (color online) 300 K neutron
powder diffraction pattern and b)-d) zoomed-in regions showing
some of the prominent supercell peaks which indicate a distortion
of the high-symmetry P$6_3/mmc$ structure where all Ni sites are
equivalent. e)-g) Supercell peaks are not observed in xray data
indicating that the distortion mainly involves displacements of
oxygen ions. Solid lines in all panels are fits to the model with
periodic arrangement of expanded and contracted NiO$_6$ octahedra
shown in Fig.\ \protect\ref{fig_structure} and peak indexing is in
the triple unit cell (space group P$6_322$). }
\end{center}
\end{figure}

Powder samples of 2H-AgNiO$_{2}$ (a polymorph \cite{Soergel05} of
the better known rhombohedral 3R-polytype \cite{Shin93}) were
prepared as described in\ Ref. \cite{Soergel05} and neutron
diffraction indicates a nearly pure hexagonal phase with less than
1\% admixture of the rhombohedral polytype. Neutron diffraction
patterns at 300 K were collected using the high-resolution
time-of-flight diffractometers OSIRIS and HRPD at the ISIS pulsed
neutron source and representative results are shown in Fig.\
\ref{fig_pattern}a), yielding the hexagonal space group
P$6_{3}/mmc$ as proposed earlier \cite{Soergel05}, with triangular
Ni, O, and Ag layers stacked as ABBBA (Ni-O-Ag-O-Ni). However the
neutron data shows additional diffraction peaks [see Fig.\
\ref{fig_pattern}b)-d)] which can be consistently indexed by an
ordering wavevector $\bm{k} _{0}=(1/3,1/3,0)$ in the original
undistorted structure, indicating a tripled unit cell in the
hexagonal plane. To test the accuracy of the indexing we have
refined the lattice parameters using only the main peaks and only
the supercell peaks (more than 20 observed) and obtained the same
values to within the experimental accuracy of $10^{-4}$\AA.
Furthermore, upon cooling to 2\ K the supercell peaks were
displaced in $Q$ following the lattice contraction but maintained
their commensurate index with respect to the main peaks. This
further confirms that the supercell peaks are due to a modulation
of the main structure rather than an extra impurity phase.

X-ray data collected using a Phillips Cu K$_{\alpha }$ powder
diffractometer [Fig\ \ref{fig_pattern}e)-g)] do not show supercell
peaks, indicating that the distortion involves mainly the light
oxygen ions. Thus we refined the neutron data in a $\sqrt{3}\times
\sqrt{3}$ hexagonal unit cell assuming only oxygen displacements.
We eliminated the space groups that were not compatible with the
observed low-temperature magnetic structure (to be described
later) dictating at least two different Ni crystallographic sites.
The highest symmetry space group in which both the structural and
magnetic data could be described is P$6_{3}22$ (\#182). Good
agreement with the data is obtained (solid line fits
\cite{fullprof} in Fig\ \ref{fig_pattern}) and the extracted unit
cell parameters are listed in Table\ \ref{tab_unitcell}. As a
further check of the space group identification we have performed
full structural optimization calculations (as described later)
starting from a lower symmetry structure (P6$_{3}$ space group)
and those invariably converged to the higher P$6_{3}22$ symmetry,
with the optimized positions agreeing well with the experimental
ones (Table \ref{tab_unitcell}) \cite{optnote}.

\begin{table}[tb]
\caption{Unit cell lattice parameters in the ideal (P$6_{3}/mmc$) and
CO (P$6_{3}22$) structures at 300 K. Internal parameters $\delta$, $\zeta$,
$\xi$, and $\epsilon$ are 0, 0.08050(5), 0 and 0.0133(2) (exp.) and
0, 0.0792, 0 and 0.0102 (calc.), respectively.
}
\label{tab_unitcell}
\begin{center}
\begin{tabular}{l|l}
\hline
P$6_3/mmc$ (no. 194) & P$6_3 2 2$ (no. 182) \\ \hline
$a_0 = 2.93919(5)$ \AA  & $a=5.0908(1)$ \AA  \\
$c= 12.2498(1)$ \AA  & $c=12.2498(1)$ \AA  \\ \hline
\begin{tabular}{lll}
Atom & Site & $(x,y,z)$ \\ \hline
Ni ~~ & 2$a$ ~~ & $(0,0,0)$ \\
~~ & ~~ &  \\
~~ & ~~ &  \\
Ag ~~ & 2$c$ ~~ & $(\frac{2}{3},\frac{1}{3},\frac{1}{4})$ \\
O ~~ & 4$f$ ~~ & $(\frac{2}{3},\frac{1}{3},\zeta)$\\[2pt]
\end{tabular}
&
\begin{tabular}{lll}
Atom & Site & $(x,y,z)$ \\ \hline
Ni1 ~~ & 2$c$ ~~ & $(\frac{1}{3},\frac{2}{3},\frac{1}{4})$ \\
Ni2 ~~ & 2$b$ ~~ & $(0,0,\frac{1}{4})$ \\
Ni3 ~~ & 2$d$ ~~ & $(\frac{1}{3},\frac{2}{3},\frac{3}{4})$ \\
Ag ~~ & 6$g$ ~~ & $(\frac{2}{3}+\delta,0,0)$ \\
O ~~ & 12$i$ ~~ & $(\frac{1}{3}+\xi,\epsilon,\frac{1}{4}+\zeta)$\\[2pt]
\end{tabular}
\\ \hline
\end{tabular}
\end{center}
\end{table}

\begin{figure}[tbph]
\begin{center}
\includegraphics[width=7.2cm,bbllx=53,bblly=247,bburx=535,
bbury=598,angle=0,clip=] {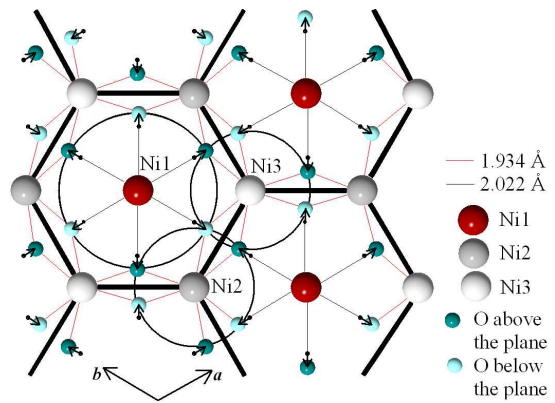}
\end{center}
\caption{(color online) Bottom NiO$_{2}$ layer showing the
expanded Ni1O$_{6}$ octahedra (large circle) surrounded by a
honeycomb network (thick lines) of contracted Ni2 and Ni3 sites
(small circles). The small balls are oxygen ions and the small
arrows indicate the displacements from the ideal structure.
%Diamond of Ni2 sites defines the in-plane unit cell.
}
\label{fig_structure}
\end{figure}

The resulting structure in one NiO$_{2}$ layer is shown in Fig.\
\ref{fig_structure}. There are three inequivalent Ni sites.
Oxygens breathe in towards Ni2 and Ni3 ions
($d_{\mathrm{Ni2-O}}=d_{\mathrm{Ni3-O}}=1.934$\ \AA ) and out from
Ni1 ($d_{\mathrm{Ni1-O}}=2.022$\ \AA ). No Jahn-Teller distortion
is present. The different Ni-O distances indicate that Ni1 is
electron rich (Ni$^{2+}$ in a CO scenario) and Ni2,3 are electron
depleted (Ni$^{3.5+}$). In the adjacent NiO$_{2}$ layer Ni1 and
Ni3 sites swap places, leading to a zigzag arrangement of the
(expanded) electron-rich Ni1 sites along $c$ \cite{note1}.

\begin{figure}[tbhp]
\begin{center}
\includegraphics[width=0.95 \linewidth,
bbllx=36,bblly=275,bburx=535,bbury=725,angle=0,clip=]
{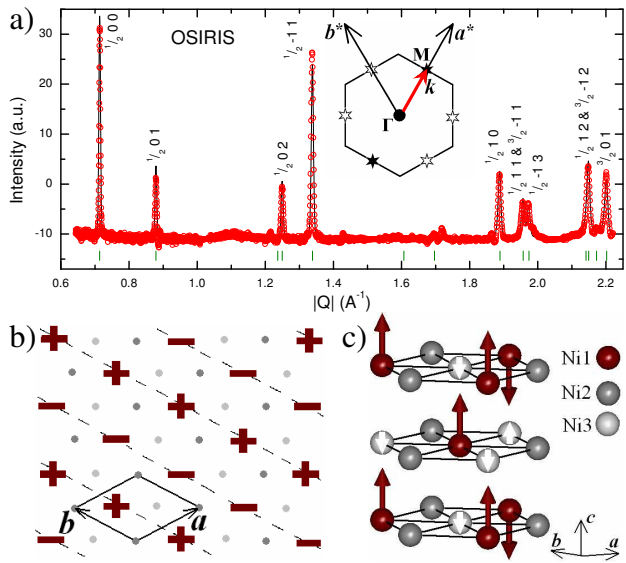}
\end{center}
%\vspace{-0.5cm}
\caption{a) (color online) Magnetic diffraction pattern
(difference 4 K - 300 K) indexed by wavevector $\bm{k}=(1/2,0,0)$.
Solid line is a fit to the spin order depicted in b): In one layer
ordered spins (Ni1) form alternating ferromagnetic rows, $\pm$
symbols indicate spin projection along the $c$-axis and gray balls
are unordered sites (Ni2 and Ni3). Diamond-shaped box is the unit
cell. c) 3D view of the magnetic order along $c$-axis.
Band-structure calculations indicate that the effective FM
inter-layer coupling between the strongly-magnetic Ni1 spins
(large arrows) occurs via AFM couplings to a very small moment at
the Ni3 site (small white arrow). Inset in a) shows the 2D
hexagonal Brillouin zone with locations of magnetic Bragg peaks
(filled stars for the stripe domain in (b), open stars for
equivalent domains rotated by $\pm60^{\circ}$).}
\label{fig_mag_pattern}
\end{figure}

Earlier susceptibility data \cite{Soergel05} showed dominant
antiferromagnetic couplings with a Curie-Weiss constant
$\theta=-107.6$ K and AFM order below T$_N$=22 K. Below this
temperature additional peaks appear in the neutron diffraction
pattern, see Fig.\ \ref{fig_mag_pattern}a). These peaks are
indexed by the ordering wavevector ${\bm k}=(1/2,0,0)$, as shown
in Fig.\ \ref {fig_mag_pattern}a) inset. The best fit was obtained
when only one Ni sublattice was ordered, either Ni1 or Ni3, with a
moment of 1.522(7) $\mu _{B}$ along the $c$-axis. As discussed
above, this must be the strongly-magnetic Ni1 (Ni$^{2+}$) with a
large available moment of $S$=1 ($2 \mu _{B})$, because Ni3 and
Ni2 are Ni$^{3.5+}$ with only a small available moment.

The observed magnetic structure is shown in Fig.\
\ref{fig_mag_pattern}b): in each layer ordered spins are arranged
in alternating FM stripes, in the adjacent layer stripes are
parallel but have an in-plane offset that follows the $c$-axis
zig-zag arrangement of the electron-rich Ni1 sites, see Fig.\
\ref{fig_mag_pattern}c). This magnetic structure implies that (a)
the exchange between nearest in-plane magnetic neighbors is AFM,
(b) there is an easy-axis anisotropy $-DS_{z}^{2}$, and (c) the
net interplanar interaction with the 3 magnetic neighbors in the
layer above (below)[see Fig.\ \ref{fig_mag_pattern}c)] is FM.

\begin{figure}[tbhp] \begin{center}
\includegraphics[width=7.5cm,bbllx=84,
bblly=70,bburx=591,bbury=375,angle=0,clip=] {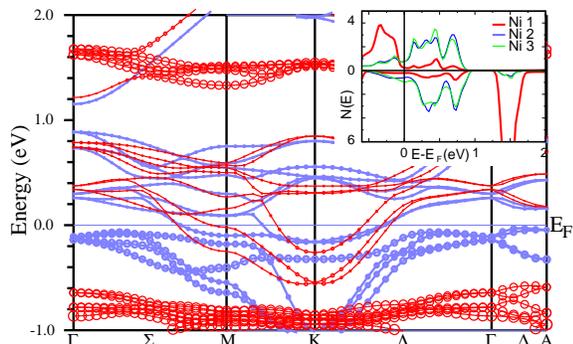}
\end{center}
\vspace{-0.7cm} \caption{(color online) Ferromagnetic band
structure of 2H-AgNiO$_2$ along symmetry directions in the
Brillouin zone. Blue/red (light/dark shades) correspond to the
up/down spin directions and circles indicate the relative weight
of the Ni1 states. Inset: Partial densities of states for the
three Ni sites in states/eV/triple cell. Top/bottom: spin majority
(up)/minority(down).} \label{fig_bands}
\end{figure}

To gain more insight into the physical origin for the observed
charge and magnetic orderings, we have performed density
functional (DFT) calculations \cite{wien,vasp}. Since
Ag$_{x}$NiO$_{2}$ is metallic and only weakly
correlated\cite{Ag2NiO2}, introducing a Hubbard $U$ in the
calculation is not necessary.  Optimizing the oxygen positions in
the $\sqrt{3}\times \sqrt{3}$ unit cell precipitates the CO and
reproduces the observed structure in Fig.\ \ref{fig_structure},
indicating that the driving force for charge disproportionation is
entirely accounted for in the LDA. The calculated total
magnetization for the three Ni ions varies between 1.3 and 1.5\
$\mu_{B}$, depending on the exact O position, and is nearly
entirely located at the expanded Ni1 site in good quantitative
accord with the neutron data. The computed ferromagnetic band
structure (Fig.\ \ref{fig_bands}) shows fully exchange-split Ni1
derived bands with the magnetic moment diminished from its formal
count of 2\ $\mu _{B}$ by hybridization. One can also see that Ni2
and Ni3 states are not visibly exchange split. The Ag $sp$ band is
entirely above the Fermi level. This electronic structure can be
visualized as a magnetic insulator formed by Ni$^{2+}$ (cf. NiO)
with a strong tendency to magnetic order, superimposed on a
Ni$^{3.5}$ metal. The latter has larger bandwidths, due to the
smaller Ni-O distance, and by itself does not satisfy the Stoner
criterion for metallic magnetism, $IN(E_{F})>1$, where $I$ is the
atomic Stoner parameter.

The natural tendency for charge order in a NiO$_2$ layer can be
explained by energetic arguments. The Hund's rule energy gain in a
metal in DFT is $M^{2}(I-N^{-1})/4,$ where $N$ is the average
density of states per spin. Without charge ordering,
$M_{1}=M_{2}=M_{3}=1 \mu _{B}$ and $ N_{1}=N_{2}=N_{3}=\bar{N}$.
After charge ordering, $M_{1}\approx 2\mu _{B}$,
$M_{2}=M_{3}\approx 0$ and $N_{1}>\bar{N}>N_{2}=N_{3}$. The CO
state is energetically favored since
$4(I-N_{1}^{-1})/4>3(I-\bar{N}^{-1})/4$. In insulators such as
NaNiO$_2$ the Hubbard repulsion would forbid this mechanism, but
in this metallic system it is well screened and cannot prevent CO
\cite{walt}. Our band-structure calculations reproduce fully the
observed 3D CO pattern, both the in-plane $\sqrt{3}\times\sqrt{3}$
order as well as the zig-zag arrangement of the electron rich
sites along $c$. We also found a metastable solution where the CO
layers form straight columns along $c$, i.e. the electron rich and
strongly magnetic Ni$^{2+}$ occupy the Ni2 site; this indicates
that the in-plane CO is very robust and independent of the
inter-layer magnetic interactions.

To study the stability of the observed magnetic structure we have
calculated ground state energies for a number of potential
structures using DFT for the full magnetic unit cell (12 formula
units and lower symmetry, C2). In addition to the observed
stripe-order with the zig-zag ferromagnetic pattern along $c$
shown in Figs.\ \ref{fig_mag_pattern}b-c) (called stripe-F), we
also found stable solutions for hypothetical ferromagnetic (FM),
A-type AFM (alternating FM layers), as well as in-plane
stripe-order but with a zig-zag AFM pattern along $c$ (stripe-A,
every other plane flipped compared to stripe-F). The results are
listed in Table\ \ref{tr}, the state found experimentally has the
lowest energy.

\begin{table}[tbp]
\caption{\label{tr} Energies of different magnetic configurations
for 2H-AgNiO$_2$. All energies are expressed per magnetic Ni1 ion
and as a difference from the calculated (and observed) magnetic
ground state.}
\begin{tabular}{|l|cccc|}
\hline & stripe-F & stripe-A & FM
& A-type AFM \\
 \hline
Energy/Ni1 (meV)  & 0 & 7 & 23 & 21\\
 \hline
\end{tabular}
\end{table}

The magnetic structure indicates that the in-plane exchange
between magnetic neighbors is AFM. This arises naturally because
Ni1$^{2+}$ sites are insulating and lack both metallic Stoner
ferromagnetism and 90$^{\circ}$ FM superexchange, but do have a
classical AFM superexchange via the Ni1-O-O-Ni1 path, with
effective hopping $ t_{eff}\sim t_{pd\sigma }^{2}t_{pp\pi
}/(E_{d}-E_{p})^{2}$. The appearance of collinear order instead of
the conventional 120$^{\circ}$ order for a triangular AFM implies
a significant easy-axis anisotropy. This is normally provided by
spin-orbit induced coupling to the crystal field, but the Ni1 ion
is close to a $t_{2g}^{6}e_{g}^{2}$ configuration with zero
orbital moment. Further experimental and theoretical studies of
the single-site anisotropy in this system are under way. The fact
that the effective interplanar Ni1-Ni1 exchange is FM is rather
surprising since there is {\em a priori} no apparent physical
mechanism for FM couplings. Indeed, the magnetic species, Ni1,
forms insulating bands and Ag is nonmagnetic and absent from the
Fermi energy, eliminating double exchange as a possibility. Ni2
and Ni3 form quasi-2D bands with small potential for sizeable RKKY
interaction, and magnetic impurities in the Ag layer that could
mediate a FM exchange have been excluded experimentally.

Performing calculations for a wider range of parameters, we find
that depending on the exact positions of Ag and O the interplanar
interaction can change amplitude and even sign, and that an
effective FM interlayer coupling can appear if the weakly-magnetic
site Ni3 has a small ordered moment (0.10-0.15 $\mu_B$). The
strongest inter-planar interaction is AFM superexchange between
the Ni3 and Ni1 directly above (it occurs via a near collinear
O-Ag-O path), whereas the inter-planar Ni1-Ni1 exchange is much
weaker. If the Ni3 moment is aligned antiparallel to the Ni1 above
it then the structure in Fig.\ \ref{fig_mag_pattern}c) has a net
energy gain as it also satisfies two out of the three in-plane AFM
Ni3-Ni1 couplings. Our diffraction data in Fig.\
\ref{fig_mag_pattern}a) cannot prove or exclude the existence of
very small moments on Ni3, but the data is consistent with a Ni3
moment of $\sim 0.1 \mu_B$ anti-aligned with the Ni1 directly
above it, as it comes out of the calculations.

To summarize, we have reported high-resolution neutron diffraction
in the orbitally-degenerate 2H-AgNiO$_2$ which observe a periodic
arrangement of expanded and contracted NiO$_6$ octahedra,
naturally explained by a three-sublattice charge order pattern on
the triangular lattice of Ni sites. We have proposed that due to
Hund's rule coupling and metallic screening of the Hubbard $U$
repulsion (and static screening due to oxygen displacements) the
CO is the favored mechanism to lift the orbital degeneracy as
opposed to the conventional Jahn-Teller distortions that occurs in
more insulating systems. An unusual magnetic order is observed at
low temperatures with only one third of sites carrying a magnetic
moment with an unexpected collinear stripe order pattern on an
antiferromagnetic triangular lattice. Our results indicate the
complex cooperative charge and magnetic order patterns that can
occur in orbitally-degenerate metallic systems on frustrated
lattices.

%To summarize, using high-resolution neutron diffraction we have
%observed a novel 2D spin and charge order pattern in the layered
%triangular magnet AgNiO$_{2}$. Charge order leads to three
%non-equivalent Ni sites, out of which only one is substantially
%magnetic. DFT calculations show that CO occurs in order to remove
%the orbital degeneracy of the formally Ni$^{3+}$ ions. Thus, the
%Jahn-Teller distortion seen in more insulating systems is replaced
%by CO, driven by Hund's rule \cite {RENIO3} and/or covalent
%\cite{walt} energy, made possible by an efficient metallic
%screening of the Hubbard $U$ repulsion. In the CO phase only
%second neighbor magnetic interactions are relevant and lead to
%strong AFM superexchange between the magnetic Ni sites located on
%an effectively tripled triangular lattice. In this frustrated
%geometry spins order in a collinear stripe pattern of alternating
%FM rows. We have proposed that the absence of the 120$^{\circ}$
%order (and other up-up-down type orders proposed for stacked
%easy-axis triangular AFM's \cite{ITAFM}) is likely due to a
%combination of easy-axis anisotropy and the interlayer couplings.
%Our calculations indicated that an effective interlayer FM
%coupling can arise indirectly through the polarization of a small
%ordered moment on one of the weakly-magnetic Ni sublattices. Our
%results illustrate the complex spin and charge order patterns that
%can occur in orbitally-degenerate metallic systems on frustrated
%lattices.

We acknowledge support from EPSRC UK.
%(grants EP/C51078X/2(EW) and
%GR/R76714/02(RC)), and a studentship from EPSRC and ILL (EMW).
%%%%%%%%%%%%%%%%%%%%%%%%%%%%%%%%%%%%%%%%%%%%%%%%%%%%%%%%%%%%%%%%%%%%%

%%%%%%%%%%%%%%%%%%%%%%%%%%%%%%%%%%%%%%%%%%%%%%%%%%%%%%%%%%%%%%%%%%%%%


\vspace{-0.75cm}
\begin{thebibliography}{99}
\vspace{-0.75cm}
%\bibitem{Vernay04} F. Vernay \emph{et al}, %, K. Penk, P. Fazekas, F. Mila,
%\prb\textbf{70}, 014428 (2004).

\bibitem{ong}
N. P. Ong and R. J. Cava, Science \textbf{305}, 52 (2004).
%Electronic frustration on a triangular lattice

\bibitem{jansen} M. Jansen and R. Hoppe, Z.
Anorg. Allg. Chem. \textbf{408}, 104 (1974).

\bibitem{NaNiO2} E. Chappel \emph{et al.},
%M. D. N\'{u}\~{n}ez-Regueiro, G. Chouteau,
%O. Isnard, and C. Darie,
Eur. Phys. J. B \textbf{17}, 615 (2000).
%M. J. Lewis \emph{et al}
%B. D. Gaulin, L. Filion, C. Kallin, A. J. Berlinsky,
%H. A. Dabkowska, Y. Qiu, and J. R. D. Copley,
%\prb \textbf{72}, 014408 (2005).

\bibitem{RENIO3} I.I. Mazin \emph{et al}, \prl \textbf{98}, 176406
(2007).

\bibitem{Soergel05} T. S\"{o}rgel and M. Jansen, Z. Anorg. Allg. Chem.
\textbf{631}, 2970 (2005); J. Solid State Chem. \textbf{180}, 8
(2007).

\bibitem{Shin93} Y.J. Shin \emph{et al.},
%, J.P. Doumerc, P. Dordor, C. Delmas, M. Pouchard,
%P. Hagenmuller,
J. Solid State Chem. \textbf{107}, 303 (1993).

\bibitem{ag2} M. Schreyer and M. Jansen, Angew. Chem. \textbf{41}, 643
(2002); U. Wedig {\em et al}, %, Peter Adler, Jürgen Nuss, Hartwig Modrow and Martin Jansen
Solid State Sci.\ \textbf{8}, 753 (2006); H. Yoshida {\em et al},
\prb \textbf{73}, 020408(R) (2006).

\bibitem{RENIO3_exp} J.A. Alonso \emph{et al}, \prl \textbf{82}, 3871
(1999).

\bibitem{fullprof} J. Rodr\'{\i}guez-Carvajal, Physica B \textbf{192}, 55
(1993).

\bibitem{optnote} Experimentally, Ni1-O bond is 4.5\%
longer than the other two Ni-O, bonds; the calculations find
3.5\%. The optimized position of O is mainly insensitive to the
position of Ag ions. The Ag ions (located between two O's in the
high symmetry structure) do not follow the O ions upon the
breathing distortion, remaining instead in a practically ideal
triangular lattice. These facts indicate that CO is an intrinsic
instability of the NiO$_2$ layer which is only loosely bound to
the close-packed Ag layer.

\bibitem{note1} Prominent supercell peaks (Fig.\ \ref{fig_pattern}b,c) such
as (011) and (212) rule out an arrangement where the expanded Ni
sites are directly above each other in adjacent layers.

\bibitem{wien} P. Blaha \emph{et al},
%, K.Schwarz, G.K.H. Madsen, D. Kvasnicka, and J. Luitz,
WIEN2K,
%, An Augmented Plane Wave + Local Orbitals Program for
%Calculating Crystal Properties, %Karlheinz Schwarz,
Techn. Universit\"{a}t Wien, Austria, 2001.
%ISBN 3950103142

\bibitem{vasp} G. Kresse and J. Furthmuller, Phys. Rev. B \textbf{54}, 11169
(1996).

\bibitem{Ag2NiO2} M. D. Johannes \emph{et al}, \prb (in press 2007).
%S. Streltsov, I. I. Mazin, D. I. Khomskii, Phys. Rev. B (in press)

\bibitem{walt} This analysis does not take into account increase on the Ni-O
covalent energy that also favors charge ordering (W.A. Harrison,
\prb \textbf{74}, 245128 (2006)).

%\bibitem{ITAFM}
%D. Blankschtein {\em et al}, \prb \textbf{29}, 5250 (1984).

\end{thebibliography}
\end{document}